\documentclass[a4paper]{spie}  

\newcommand{\Hitomi}{\textit{Hitomi}} 
\newcommand{\Suzaku}{\textit{Suzaku}} 
 
\usepackage{amsmath,amsfonts,amssymb}
\usepackage[dvipdfmx]{graphicx}
\usepackage[colorlinks=true, allcolors=blue]{hyperref}
\usepackage{aas_macros}

\title{Xtend, the Soft X-ray Imaging Telescope for the X-ray Imaging and Spectroscopy Mission (XRISM)}

\author[a,b]{Koji Mori}
\affil[a]{Faculty of Engineering, University of Miyazaki, 1-1 Gakuen Kibanadai Nishi, Miyazaki, Miyazaki 889-2192, Japan}
\affil[b]{Japan Aerospace Exploration Agency, Institute of Space and Astronautical Science, 3-1-1 Yoshino-dai, Chuo-ku, Sagamihara, Kanagawa 252-5210, Japan}

\author[c]{Hiroshi Tomida}
\affil[c]{Japan Aerospace Exploration Agency, Institute of Space and Astronautical Science, 2-1-1, Sengen, Tsukuba, Ibaraki 305-8505, Japan}
\author[d,b]{Hiroshi Nakajima}
\affil[d]{College of Science and Engineering, Kanto Gakuinn University, Kanazawa-ku, Yokohama, Kanagawa 236-8501, Japan}

\author[e]{Takashi Okajima}
\affil[e]{NASA's Goddard Space Flight Center, Greenbelt, MD 20771, USA}

\author[f,g]{Hirofumi Noda}
\affil[f]{Department of Earth and Space Science, Osaka University, 1-1 Machikaneyama-cho, Toyonaka, Osaka 560-0043, Japan}
\affil[g]{Forefront Research Center, Osaka University, 1-1 Machikaneyama-cho, Toyonaka, Osaka 560-0043, Japan}
\author[h]{Takaaki Tanaka}
\affil[h]{Department of Physics, Konan University, 8-9-1 Okamoto, Higashinada, Kobe,
Hyogo 658-8501}
\author[i]{Hiroyuki Uchida}
\affil[i]{Department of Physics, Kyoto University, Kitashirakawa Oiwake-cho,Sakyo-ku, Kyoto, Kyoto 606-8502, Japan}

\author[d]{Kouichi Hagino}
\author[j]{Shogo Benjamin Kobayashi}
\affil[j]{Department of Physics, Faculty of Science, Tokyo University of Science, Kagurazaka, Shinjuku-ku, Tokyo 162-0815, Japan}
\author[h]{Hiromasa Suzuki}
\author[c]{Tessei Yoshida}

\author[k]{Hiroshi Murakami}
\affil[k]{Faculty of Liberal Arts, Tohoku Gakuin University, 2-1-1 Tenjinzawa, Izumi-ku, Sendai, Miyagi 981-3193, Japan}

\author[l]{Hideki Uchiyama}
\author[m]{Masayoshi Nobukawa}
\author[n]{Kumiko Nobukawa}
\affil[l]{Science Education, Faculty of Education, Shizuoka University, Suruga-ku, Shizuoka, Shizuoka 422-8529, Japan}
\affil[m]{Faculty of Education, Nara University of Education, Nara, Nara 630-8528, Japan}
\affil[n]{Department of Physics, Kindai University, 3-4-1 Kowakae, Higashi-Osaka, Osaka 577-8502, Japan}

\author[b]{Tomokage Yoneyama}

\author[f,g]{Hironori Matsumoto}
\author[i]{Takeshi Tsuru}

\author[a]{Makoto Yamauchi}
\author[a]{Isamu Hatsukade}

\author[b]{Manabu Ishida}
\author[b]{Yoshitomo Maeda}
\author[e,o]{Takayuki Hayashi}
\author[e,o]{Keisuke Tamura} 
\author[e,o]{Rozenn Boissay-Malaquin} 
\author[p]{Toshiki Sato}

\affil[o]{Center for Space Science and Technology, University of Maryland, Baltimore County (UMBC), Baltimore, MD 21250, USA}
\affil[p]{Department of Physics, Rikkyo University, 3-34-1 Nishi Ikebukuro, Toshima-ku, Tokyo 171-8501, Japan}

\author[q]{Junko Hiraga}
\author[r]{Takayoshi Kohmura}
\author[s]{Kazutaka Yamaoka}
\affil[q]{Department of Physics, Kwansei Gakuin University, 2-2 Gakuen, Sanda, Hyogo 669-1337, Japan}
\affil[r]{Department of Physics, Faculty of Science and Technology, Tokyo University of Science, 2641 Yamazaki, Noda, Chiba 270-8510, Japan}
\affil[s]{Department of Physics, Nagoya University, Chikusa-ku, Nagoya,	Aichi 464-8602, Japan}

\author[b,t]{Tadayasu Dotani}
\author[b]{Masanobu Ozaki}
\author[f]{Hiroshi Tsunemi}
\affil[t]{Department of Space and Astronautical Science, School of Physical Sciences,
	SOKENDAI (The Graduate University for Advanced Studies),
	3-1-1 Yoshino-dai, Chuou-Ku, Sagamihara, Kanagawa 252-5210, Japan}

\author[a]{Yoshiaki Kanemaru}
\author[a]{Jin Sato}
\author[a]{Toshiyuki Takaki}
\author[a]{Yuta Terada}
\author[a]{Keitaro Miyazaki}
\author[a]{Kohei Kusunoki}
\author[a]{Yoshinori Otsuka}
\author[a]{Haruhiko Yokosu}
\author[a]{Wakana Yonemaru}

\author[d]{Yoh Asahina}

\author[f]{Kazunori Asakura}
\author[f]{Marina Yoshimoto}
\author[f]{Yuichi Ode}
\author[f]{Junya Sato}
\author[f]{Tomohiro Hakamata}
\author[f]{Mio Aoyagi}

\author[n]{Yuma Aoki}

\author[r]{Shun Tsunomachi}
\author[r]{Toshiki Doi}
\author[r]{Daiki Aoki}
\author[r]{Kaito Fujisawa}
\author[r]{Masatoshi Kitajima}

\author[f]{Kiyoshi Hayashida}

\authorinfo{Further author information: (Send correspondence to K.M.)\\K.M.: E-mail: mori@astro.miyazaki-u.ac.jp, Telephone: 81 985 58 7371}
 
\begin{document} 
\maketitle

\begin{abstract}
Xtend is a soft X-ray imaging telescope developed for the X-Ray Imaging and
Spectroscopy Mission (XRISM). XRISM is scheduled to be launched in the Japanese
fiscal year 2022. Xtend consists of the Soft X-ray Imager (SXI), an X-ray CCD
camera, and the X-ray Mirror Assembly (XMA), a thin-foil-nested conically
approximated Wolter-I optics. The SXI uses the P-channel, back-illuminated type CCD
with an imaging area size of 31 mm on a side. The four CCD chips are arranged in a
2$\times$2 grid and can be cooled down to $-120$ $^{\circ}$C with a single-stage
Stirling cooler. The XMA nests thin aluminum foils coated with gold in a confocal
way with an outer diameter of 45~cm. A pre-collimator is installed in front of the
X-ray mirror for the reduction of the stray light. Combining the SXI and XMA with a
focal length of 5.6m, a field of view of $38^{\prime}\times38^{\prime}$ over the
energy range from 0.4 to 13 keV is realized. We have completed the fabrication of
the flight model of both SXI and XMA. The performance verification has been
successfully conducted in a series of sub-system level tests. We also carried out
on-ground calibration measurements and the data analysis is ongoing.
\end{abstract}


\section{INTRODUCTION}
\label{sec:intro}  

\begin{figure} [ht]
 \begin{minipage}{0.49\hsize}
  \centering \includegraphics[width=\textwidth]{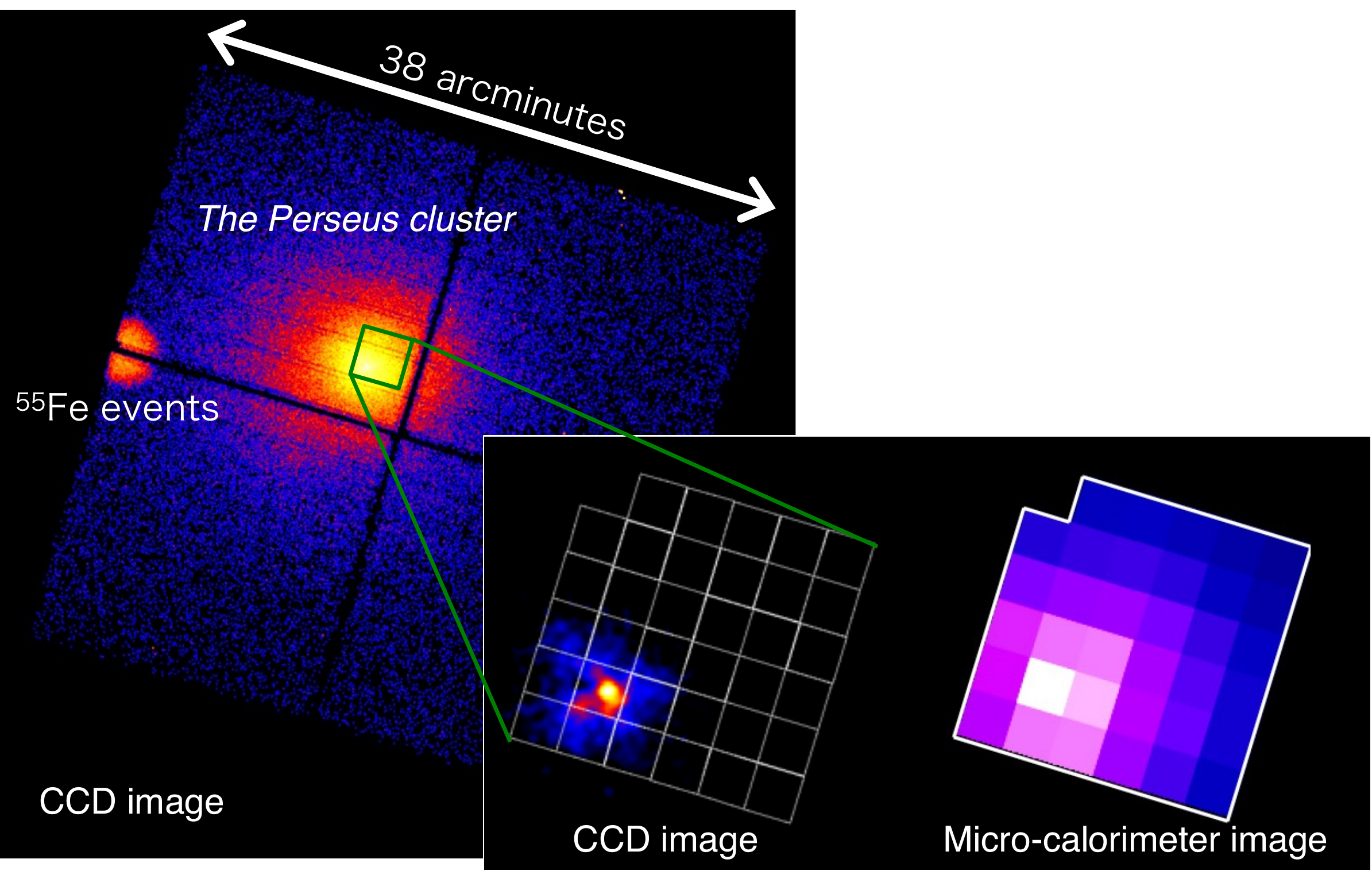}
  \caption[Perseus]{\label{fig:Perseus} X-ray images of the Perseus Cluster (left)
  and its central radio galaxy NGC~1275 (right) taken with
  \Hitomi\cite{2018PASJ...70...21N}. The intensity scales of the two CCD images
  (left and right) are different to show the fainter entire cluster view and its
  bright central object. White grids superimposed on the CCD image of NGC~1275
  indicate the micro-calorimeter pixel array.}
 \end{minipage}
 \hspace{0.02\hsize}
  \begin{minipage}{0.49\hsize}
   \centering \includegraphics[width=\textwidth]{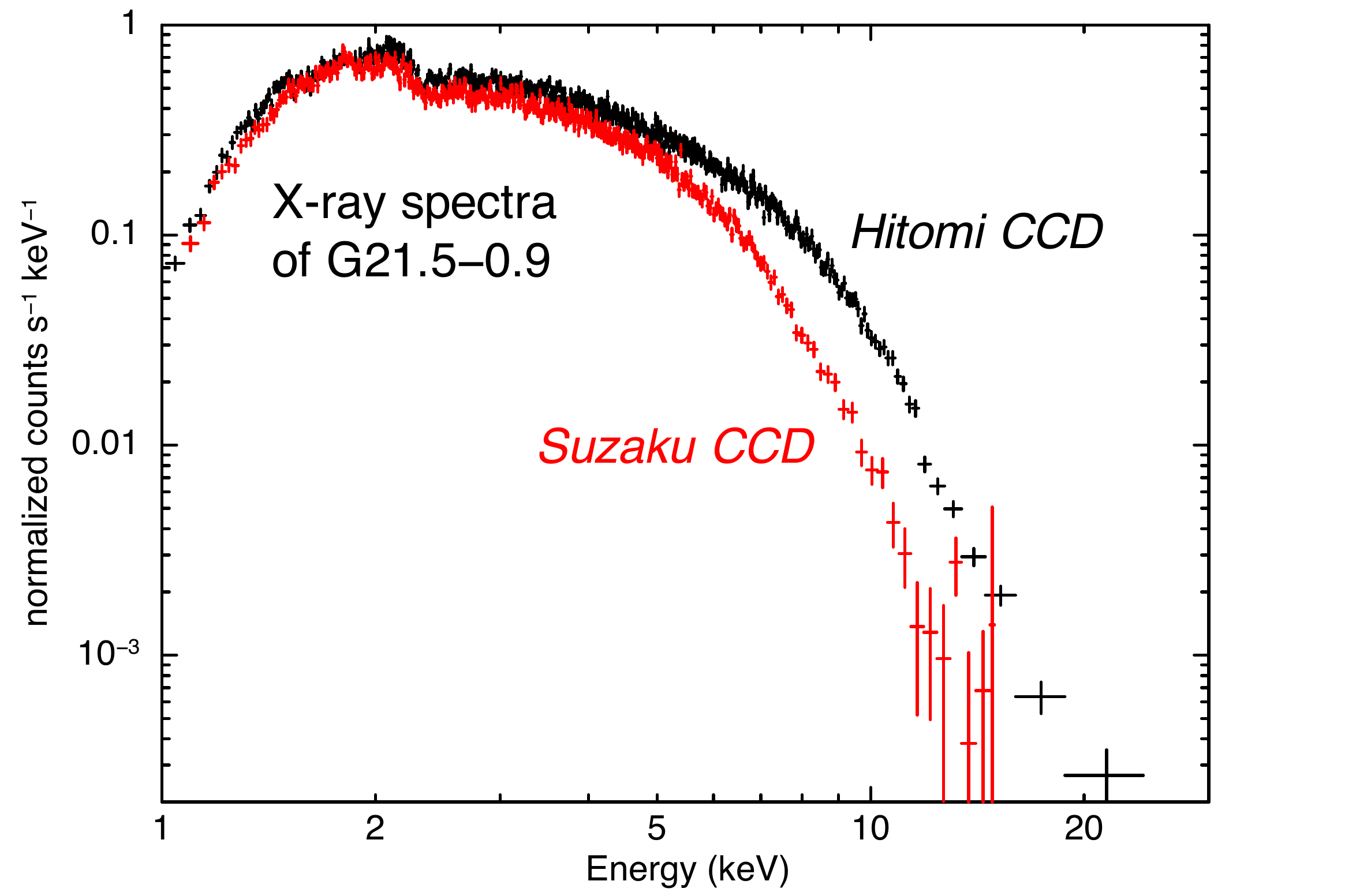}
   \caption[G21.5]{\label{fig:G21.5} X-ray spectra of G21.5$-$0.9 taken with
   \Hitomi\cite{2018PASJ...70...38A} (black) and \Suzaku\cite{2011A&A...525A..25T}
   (red) CCD cameras. \Suzaku\ carried four CCD cameras and the \Suzaku\ spectrum
   was taken from the one using a back-illuminated type CCD.}
 \end{minipage}
\end{figure}

X-Ray Imaging and Spectroscopy Mission (XRISM) is the Japan's seventh X-ray
astronomy mission and aimed at recovering one of the key scientific goals of the
previous \Hitomi\ mission, focusing only on the imaging and spectroscopy in the soft
X-ray band\cite{2018SPIE10699E..22T, 2020SPIE11444E..22T}. XRISM is developed in the
international collaboration of JAXA, NASA, ESA, and other institutes, and its launch
is scheduled in the Japanese fiscal year 2022 on a JAXA H-$\rm I\hspace{-.01em}I$A
rocket. XRISM carries two telescopes, Xtend\cite{2018SPIE10699E..23H} and
Resolve\cite{2018JLTP..193..991I}. Xtend is a soft X-ray ``imaging'' telescope,
while Resolve is a soft X-ray ``spectroscopy'' telescope. Each of them consists of
an X-ray Mirror Assembly (XMA)\cite{2018JATIS...4a1213I}, a thin-foil-nested
conically approximated Wolter-I optics, and a focal plane detector with a focal
length of 5.6~m. The focal plane detector of Xtend is the Soft X-ray Imager (SXI),
an X-ray CCD camera\cite{2020SPIE11444E..23N}, while that of Resolve is an X-ray
micro-calorimeter. Fig.~\ref{fig:Perseus} shows X-ray images of the Perseus Cluster
and its central radio galaxy NGC~1275 taken with the \Hitomi\ CCD camera and
micro-calorimeter. Since the instrument design of XRISM are basically the same as
those flown on \Hitomi, this figure demonstrates the imaging capability of
XRISM. The large field of view (FoV) of $38^{\prime}\times38^{\prime}$ realized by
Xtend encompasses the $3^{\prime}\times3^{\prime}$ FoV of Resolve in its center. The
finer pixel size of $1.\!\!^{\prime\prime}74$ of Xtend compared to that of
$30^{\prime\prime}$ of Resolve well over-samples the point spread function of the
XMA with an angular resolution of $1.\!\!^{\prime}7$ and provides more detailed
spatial information. On the other hand, Resolve realizes an unprecedented energy
resolution of 7~eV in FWHM and will revolutionize our understanding of the X-ray
Universe. Xtend and Resolve have their own characteristics in imaging and
spectroscopy, respectively, and play complementary roles in XRISM.

Xtend's SXI employs a new type of P-channel back-illuminated type CCD with a thick
depletion layer of 200~$\mu$m\cite{ozawa_2007, Takagi_2007, Ueda_2011}, following
the \Hitomi\ CCD camera\cite{2018JATIS...4a1211T}. Japan's fifth X-ray astronomy
mission, \Suzaku\, carried four CCD cameras, one of which used a back-illuminated
type CCD with a depletion layer thickness of about
40~$\mu$m\cite{2007PASJ...59S..23K}.  Fig.~\ref{fig:G21.5} shows X-ray spectra of
G21.5$-$0.9 taken with the \Hitomi\ and the \Suzaku\ CCD
cameras\cite{2018PASJ...70...38A, 2011A&A...525A..25T}, indicating the power of the
thicker depletion layer of the CCD employed by the \Hitomi\ X-ray CCD camera and
also the Xtend's SXI. The thick depletion layer is effective also in the reduction
of the non X-ray background (NXB) level above
6~keV\cite{2018PASJ...70...21N}. Xtend's XMA and \Hitomi\ X-ray mirror follow the
concept of the \Suzaku\ X-ray mirror\cite{2007PASJ...59S...9S}, nesting thin
aluminum foils coated with gold with a pre-collimator for the reduction of the stray
light. However, XMA's enlargement in the outer diameter from 40~cm to 45~cm with the
longer focal length leads to the increase of the effective area by a factor of
$\sim$1.5 \cite{2014SPIE.9144E..58I}. The high detection efficiency and low NXB
level of the SXI and the large effective area of the XMA, in addition to the
imagining capability described above, characterize the Xtend as a science instrument
onboard XRISM\cite{2018SPIE10699E..23H, 2020SPIE11444E..23N}.

\begin{figure} [ht]
  \centering \includegraphics[width=0.6\textwidth]{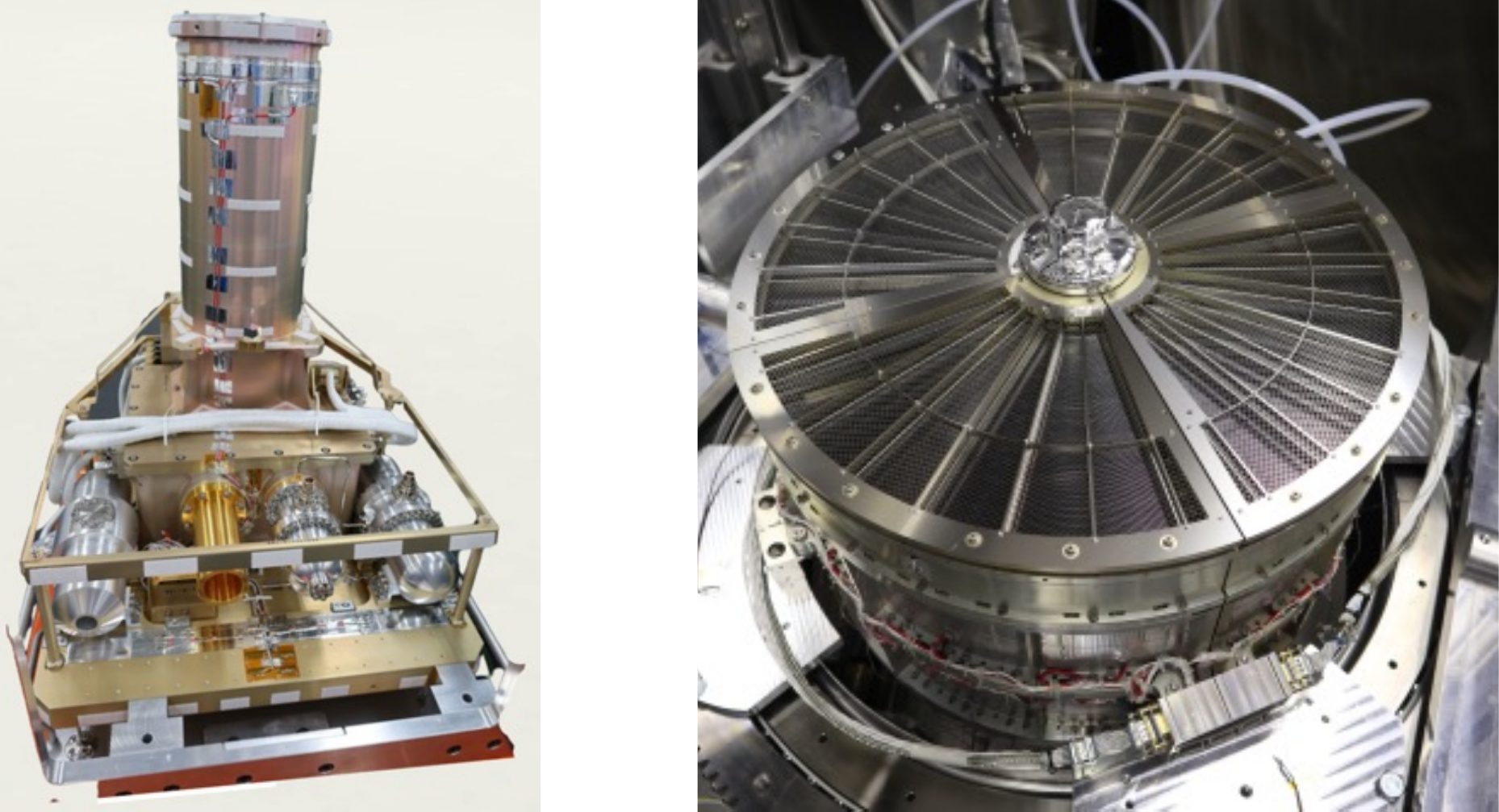}
  \caption[Perseus]{\label{fig:photo} Photographs of the flight models of the SXI
  camera body (left) and XMA (right).}
\end{figure}

We have completed the fabrication of the flight models of both SXI and XMA as shown
in Fig.~\ref{fig:photo}.  The performance verification has been successfully
conducted in a series of sub-system level tests. We also carried out on-ground
calibration measurements. In this paper, we report the current status of Xtend and
describe the performances of the sub-system tests and calibration measurements,
focusing on the SXI. The calibration studies of the XMA are presented
elsewhere\cite{Boissay-MalaquinSPIE2022, tamuraSPIE2022, hayashiSPIE2022}. CCD
spectra shown below are all made with grade 0, 2, 3, 4, and 6 events unless otherwise
indicated. The confidence level of uncertainties attached is 68\% throughout this
paper.  

\section{Xtend OVERVIEW}
\label{sec:overview}

\begin{figure} [ht]
  \centering \includegraphics[width=0.6\textwidth]{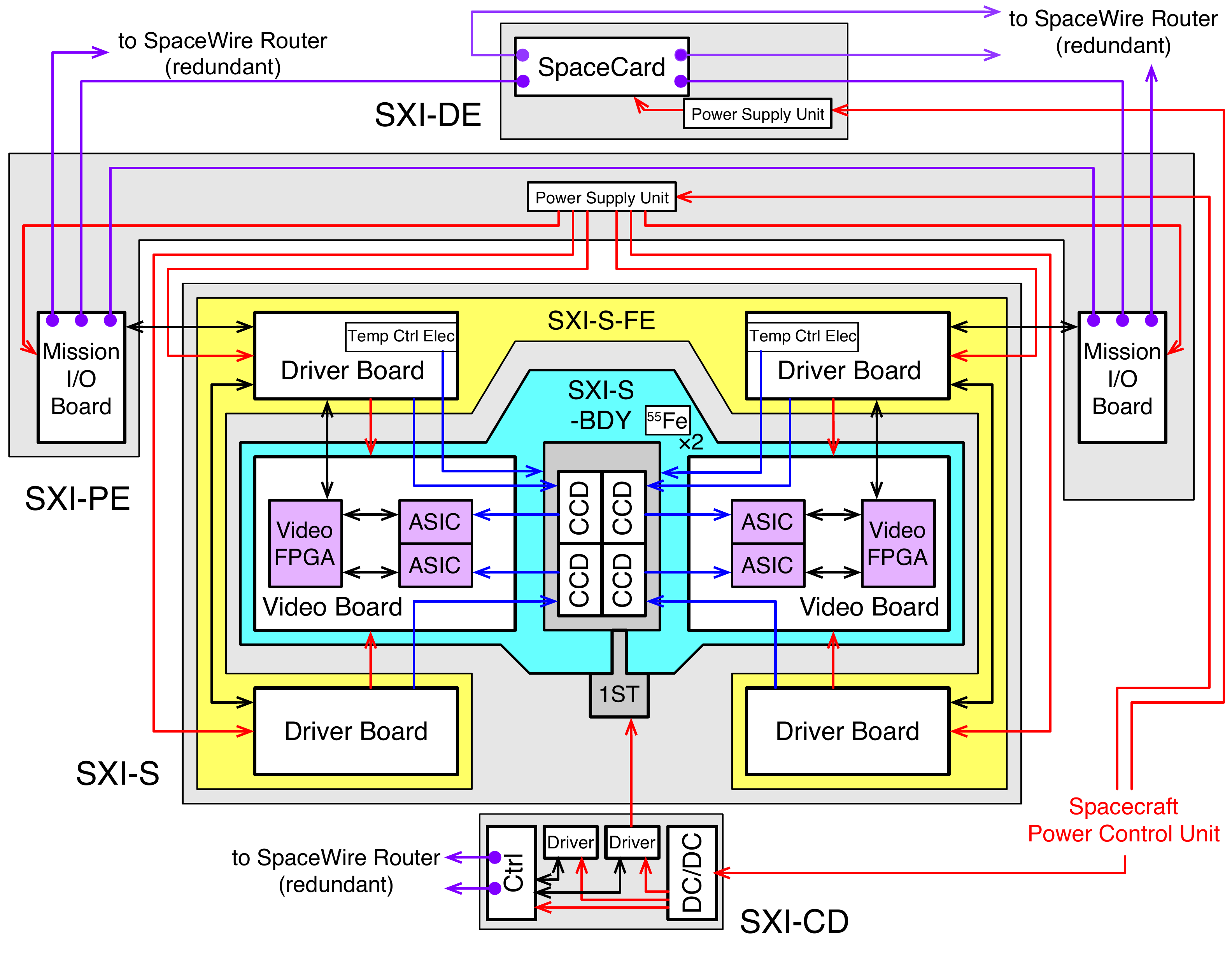}
  \caption[Perseus]{\label{fig:diagram} Block diagram of the SXI system. Red, blue,
 black, and purple lines indicate power supply, analog signal, digital signal, and
 SpaceWire connections, respectively.}
\end{figure}

The overall designs of the SXI and XMA are basically identical to those of the
\Hitomi\ X-ray CCD camera and X-ray mirror, respectively. Fig.~\ref{fig:diagram}
shows the block diagram of the SXI system.  The SXI consists of four major
components: SXI-S, SXI-PE, SXI-DE, and SXI-CD. The detailed descriptions of each
component and the changes in the CCD camera system from \Hitomi\ to XRISM are given
in Ref.\citenum{2018JATIS...4a1211T} and \citenum{2020SPIE11444E..23N},
respectively. We here provide a short summary of the SXI system. SXI-S is a sensor
part and consists of a camera body, front-end electronics, and a single-stage
Stirling cooler (1ST in Fig.~\ref{fig:diagram}). Four CCDs, arranged in a 2$\times$2
grid, are installed in the camera body with video boards. Although the SXI system is
designed to be capable of cooling CCDs down to $-120 ^{\circ}$C with the Stirling
cooler, we plan to operate the SXI with a CCD temperature of $-110 ^{\circ}$C at
least in the first years after the launch. Two $^{55}$Fe radioisotopes are also
installed for calibration purpose. SXI-PE is the pixel processing electronics which
processes the CCD signals from the video board and provide the timing signals of the
CCD clocks to the driver board. Each of the two mission I/O boards equipped with
FPGAs is responsible for a pair of two CCDs. SXI-DE is the digital electronics which
further processes the CCD data based on the information processed by SXI-PE and
compiles X-ray event data. SXI-DE also controls the entire SXI system except for the
Stirling cooler, which is operated by the cooler driver, SXI-CD.

\begin{table}[ht]
 \caption{Specifications and nominal operation parameters of the SXI CCD} 
 \label{tab:CCD-spec}
 \begin{center}       
  \begin{tabular}{|l|l|l|}
   \hline
   CCD Specification & Architecture & Frame transfer  \\
    & Imaging area size & 30.720 mm $\times$ 30.720 mm \\
    & Pixel format (physical/logical) & 1280 $\times$ 1280 / 640 $\times$ 640 \\
    & Pixel size (physical/logical) & 24$\mu$m $\times$ 24$\mu$m / 48$\mu$m $\times$ 48$\mu$m \\
    & Depletion layer thickness & 200 $\mu$m \\
    & Incident surface layer (back side) & 100 nm $+$ 100 nm thick Aluminum coat \\
    & Readout nodes (equipped/used)& 4 / 2 \\
   \hline
   Operation parameters & Frame cycle & 4 seconds  \\
   & On-chip binning & 2$\times$2  \\
   & Charge injection & every 160 physical rows \\
   \hline
  \end{tabular}
 \end{center}
\end{table}

The CCDs were manufactured by Hamamatsu Photonics K.K.\ as in the case of
\Hitomi\cite{2018JATIS...4a1211T}. Tab.~\ref{tab:CCD-spec} summarizes the
specifications of the XRISM CCD, Pch-NeXT4A. Please note that physical and logical
values refer to those before and after on-chip 2$\times$2 binning,
respectively. Figures and plots in the following sections are all made in the unit
of the logical value. One CCD has four readout nodes, two of which are used to
simultaneously read out two halves of the imaging area. Therefore, two ``segments''
per CCD are taken in a frame cycle. Although most items shown in
Tab.~\ref{tab:CCD-spec} are the same as those of the \Hitomi\ CCD, Pch-NeXT4, we
introduced two major changes on the CCD structure\cite{2020SPIE11444E..23N}. One is
to add another aluminum layer on the top of the originally existing aluminum layer,
resulting in a 100~nm $+$ 100~nm thick aluminum coat, and to insert an extra
aluminum layer under the electrode layer for the sake of the secure reduction of
visible and infrared light leaks\cite{UCHIDA2020164374}. The other is to employ a
notch structure, which is a narrow implant in the CCD channel confining a charge
packet to a fraction of the pixel volume to reduce the charge transfer inefficiency
(CTI)\cite{2019JInst..14C4003K}.  We also changed the screening criteria of flight
candidate CCDs before shipment from the manufacturer\cite{YONEYAMA2021164676} and
successfully excluded ones with the CTI anomaly observed in the \Hitomi\
CCDs\cite{2018JATIS...4a1211T}.

\begin{figure} [ht]
 \begin{minipage}{0.49\hsize}
  \centering \includegraphics[width=\textwidth]{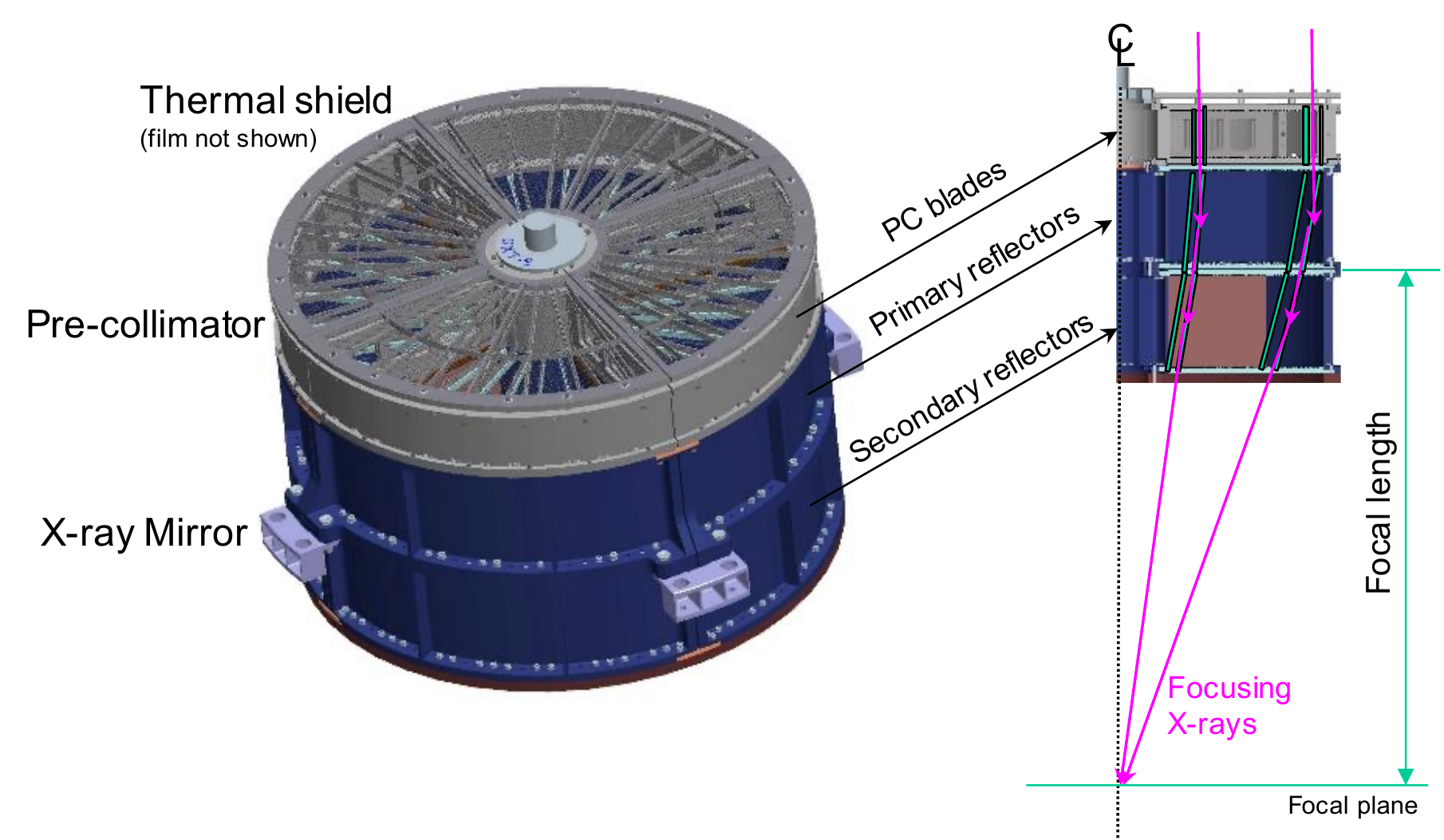}
  \caption[]{\label{fig:XMA} Schematic drawing of the XMA.}
 \end{minipage}
 \hspace{0.01\hsize}
  \begin{minipage}{0.49\hsize}
  \centering \includegraphics[width=\textwidth]{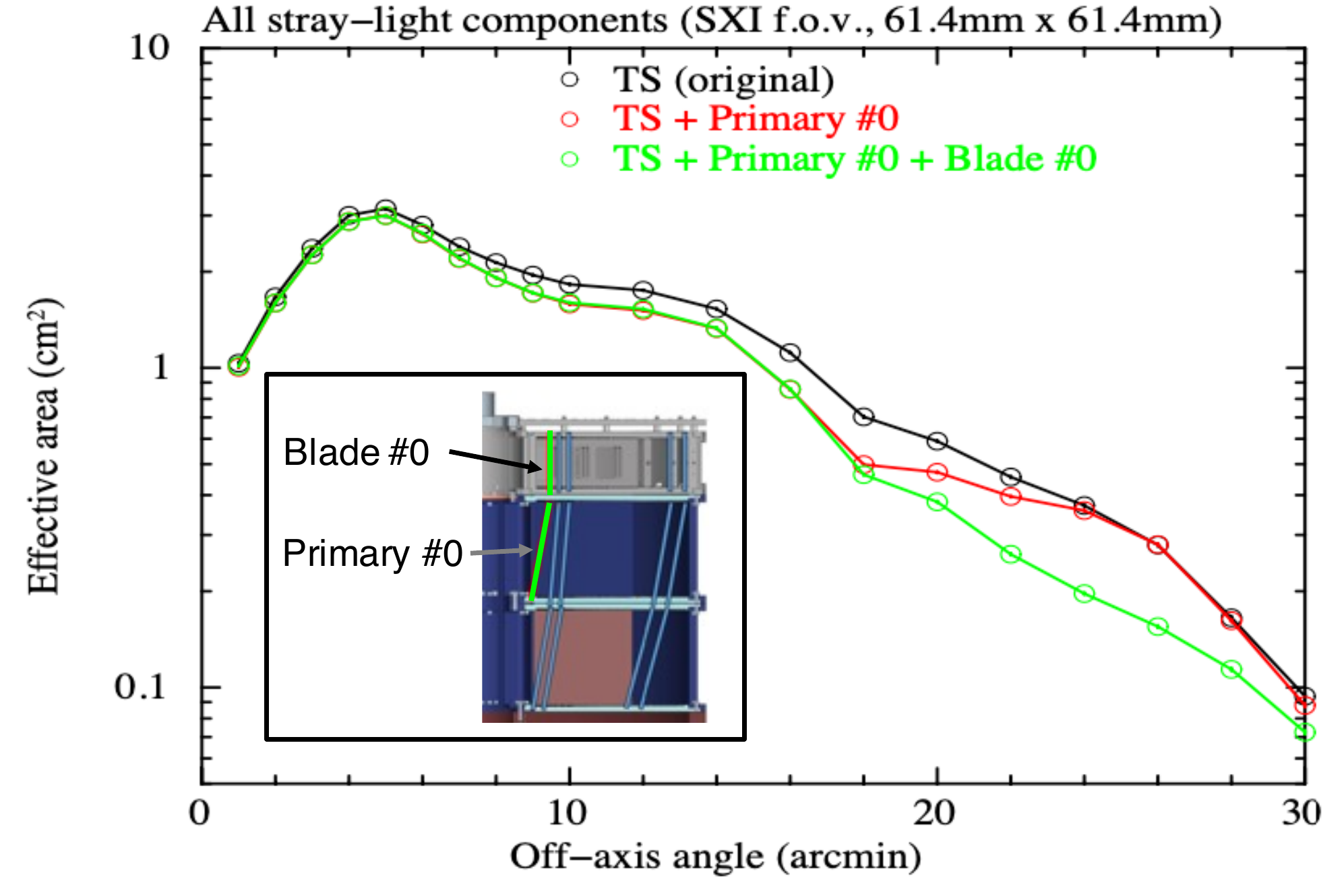}
  \caption[]{\label{fig:XMA-changePoint} Effective area curves as a function of
  off-axis angle for the stray light component at 1~keV in three cases: no change
  from \Hitomi\ (black), adding Primary \#0 (red), and adding Primary \#0 and Blade
  \#0 (green). The inset shows where Primary \#0 and Blade \#0 are inserted. }
  \end{minipage}
\end{figure}

Fig.~\ref{fig:XMA} shows schematic drawings of the XMA. The XMA conically
approximates Wolter-I grazing incidence optics. The effective diameter ranges from
116 to 450~mm. Each of the primary and secondary stages nests 203 reflectors. A
reflector consists of an aluminum substrate with epoxy-replicated gold smooth
surface for enhanced X-ray reflection. The thickness and height are 0.15--0.3~mm and
101.6~mm, respectively. The pre-collimator, placed in front of the primary stage for
the reduction of the stray light, contains co-axially nested cylindrical aluminum
blades arranged right above reflectors. A thermal shield is attached on the top of
the pre-collimator to stabilize the thermal condition of the XMA. Although this
design follows that of the \Hitomi\ X-ray mirror, a change to reduce the amount of
the stray light at large off-axis angles is
introduced. Fig.~\ref{fig:XMA-changePoint} shows the idea of this change and its
effect. In the \Hitomi\ X-ray mirror design, the innermost reflector in the second
stage is exposed to directly incoming X-rays, which could result in the stray light
at large off-axis angles. Inserting an extra blade in the pre-collimator and an
extra reflector in the primary stage, blade \#0 and primary \#0 in
Fig.~\ref{fig:XMA-changePoint}, eliminates the light path and effectively reduces
the amount of the stray light. It also helps to reduce the possibility of the MMOD
focusing.

\section{SXI Performance verification in sub-system tests}

\begin{figure} [htbp]
  \centering \includegraphics[width=\textwidth]{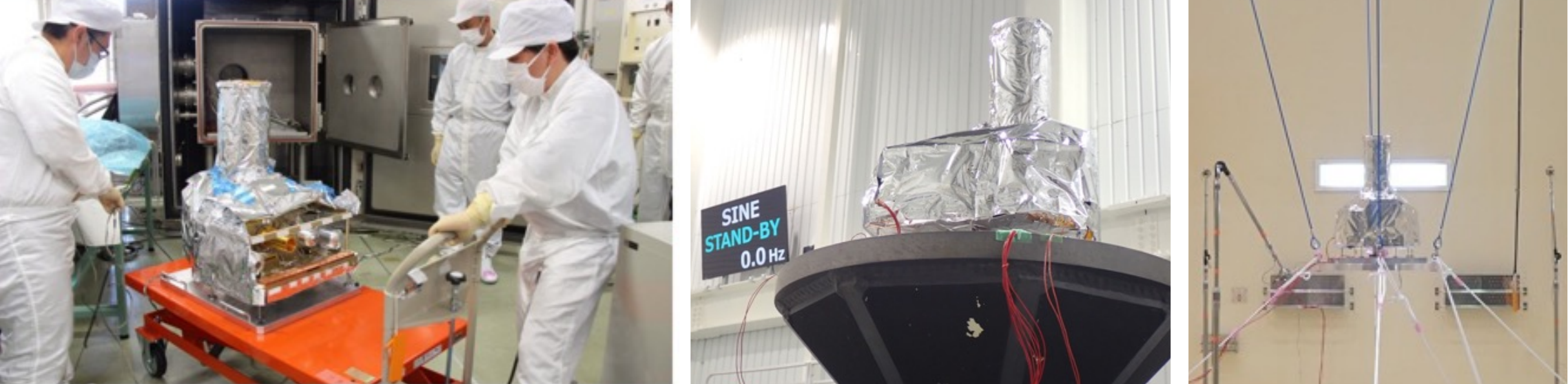}
  \caption[Perseus]{\label{fig:tests} Photographs of the SXI thermal vacuum,
  vibration, and acoustic tests from left to right. These photographs are available
  at the XRISM social media sites.}
\end{figure}

As of writing this paper, the SXI is installed into the spacecraft body and
undergoes functional tests in a course of the spacecraft checkout. Before the
installation, we successfully conducted a series of performance verification tests
in sub-system level, including the thermal vacuum, vibration, and acoustic tests, as
shown in Fig.~\ref{fig:tests}. In the SXI thermal vacuum test held in October 2021,
we verified X-ray imaging and spectroscopic performances in the full flight
configuration for the first time. Since the SXI camera body is not vacuum-tight, the
thermal vacuum test, in which the whole camera body is put into a large vacuum
chamber as shown in Fig.~\ref{fig:tests} left, was the only occasion for us to
verify its performance in the full flight configuration.

\begin{figure} [htbp]
 \begin{minipage}{0.49\hsize}
  \centering \includegraphics[width=0.8\textwidth]{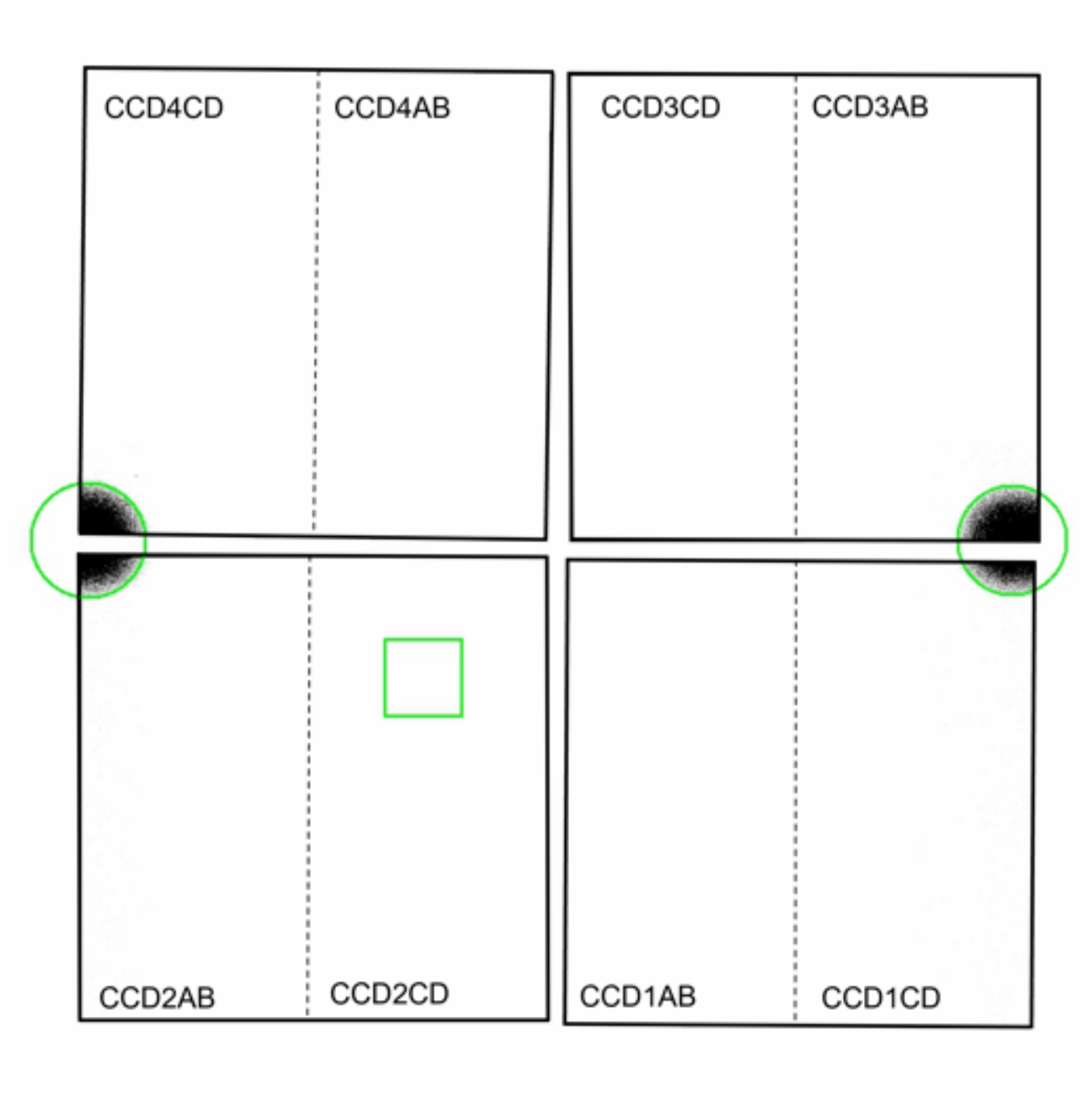}
 \end{minipage}
 \hspace{0.02\hsize}
  \begin{minipage}{0.49\hsize}
   \centering
   \includegraphics[width=\textwidth]{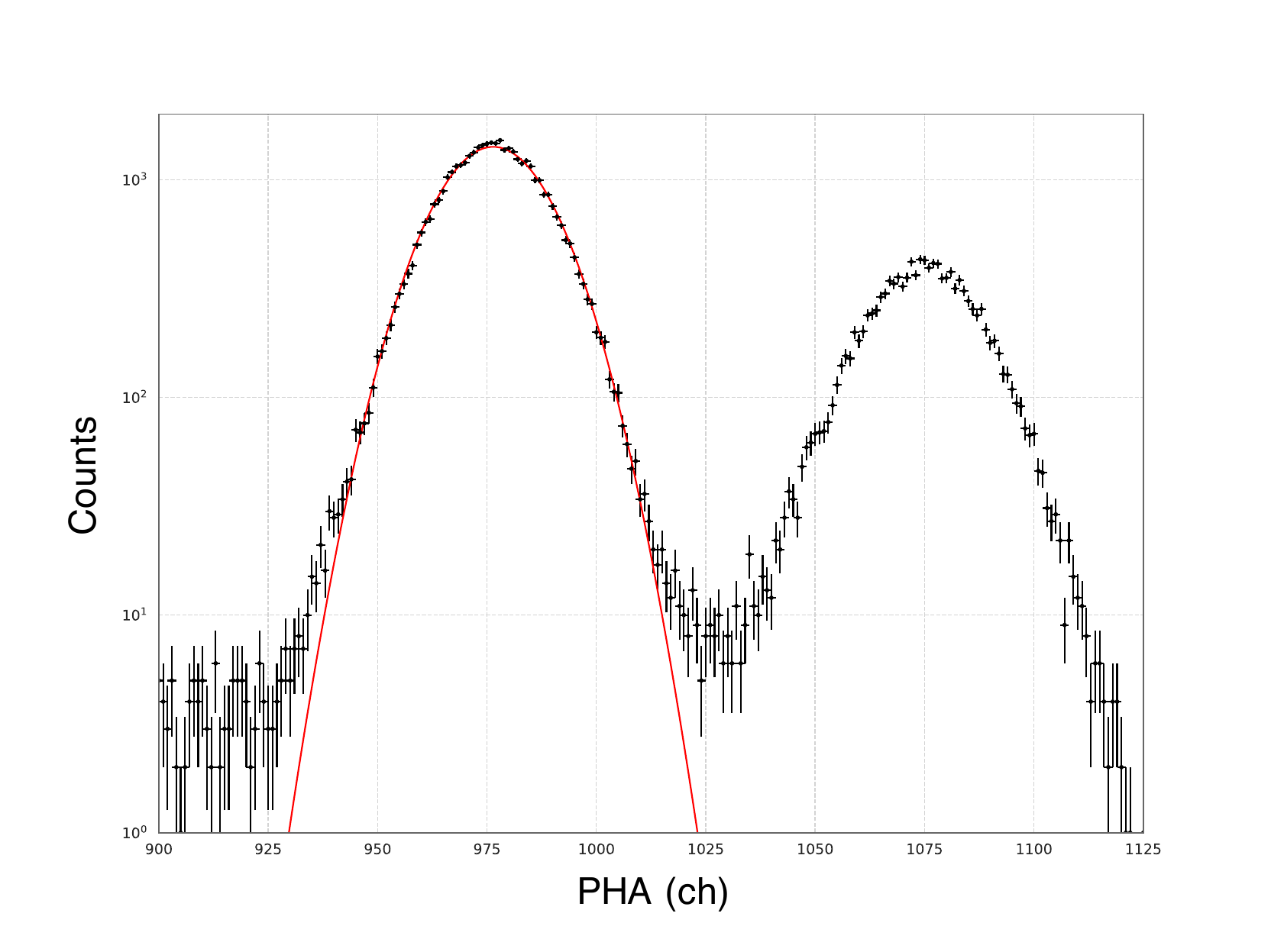}
 \end{minipage}
  \caption[Perseus]{\label{fig:TV-img-spectrum} (left) Count map of X-ray events
 taken in the thermal vacuum test in a look-down view. Green circles encompass
 irradiated regions from the $^{55}$Fe calibration sources. Black boxes and a green
 box indicate imaging areas of the four CCDs and that of the X-ray
 micro-calorimeter, respectively. Dashed lines correspond to boundaries of the CCD
 segments. Names of the CCDs (1--4) and segments (AB or CD) are
 superimposed. (right) Spectrum of X-ray events taken in the thermal vacuum
 test. The spectrum is made using the data taken in the segment CCD2AB. The red
 curve is the best-fit single Gaussian to the Mn-K$\alpha$ line.}
\end{figure}

\begin{figure} [htbp]
  \centering \includegraphics[angle=270,
  width=0.7\textwidth]{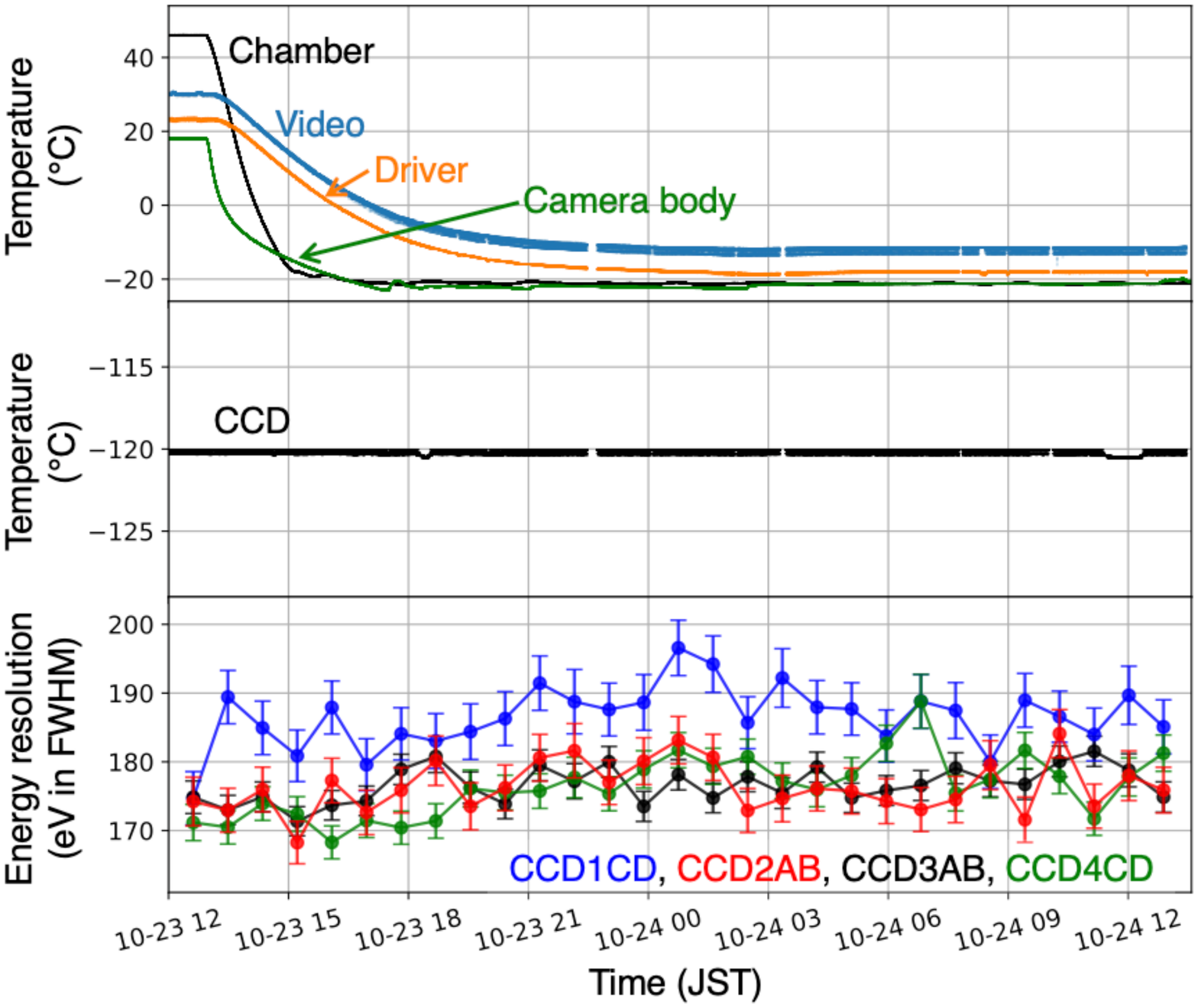}
  \caption[Perseus]{\label{fig:TV-history} Time histories of temperatures of
  the chamber, video board, driver board, and camera body (top), CCD temperature (middle),
  and energy resolutions measured in the segments of CCD1CD, CCD2AB, CCD3AB, and
 CCD4CD (bottom).}
\end{figure}

Fig.~\ref{fig:TV-img-spectrum} shows a count map and spectrum of X-ray events from
the $^{55}$Fe calibration source taken in the thermal vacuum test. From the count
map, we confirmed that the count rate, irradiated location, and region size are as
designed. The spectrum is made with events taken in the segment CCD2AB, and CTI
corrections\cite{2014NIMPA.765..269N, KANEMARU2020164646} are not applied. The
energy resolution of this spectrum is 177.5$\pm$0.6~eV in FWHM and as expected from
on-ground calibrations held in advance, which is discussed in the next section.

Fig.~\ref{fig:TV-history} shows the time history of energy resolutions measured in
the four segments irradiated with X-rays from the calibration sources as well as
that of temperatures of the chamber, camera body, video, driver boards, and CCD. The
spectrum in Fig.~\ref{fig:TV-img-spectrum} is an integrated one during the entire
time period shown in Fig.~\ref{fig:TV-history}. In this thermal vacuum test, the
temperature of the camera body was forced to swing from the hottest case
($+17~^{\circ}$C) to the coldest case ($-20~^{\circ}$C) defined by the system,
during which the SXI system successfully kept the CCD temperature constant at
$-120~^{\circ}$C. The temperatures of the video and driver boards are not under
control in the SXI system. It was verified that the energy resolutions of the CCDs
are stable within the statistical uncertainties and insensitive to external thermal
environment variations.

\section{SXI On-ground Calibration}

\begin{figure} [htbp]
 \begin{minipage}{0.49\hsize}
   \centering \includegraphics[width=\textwidth]{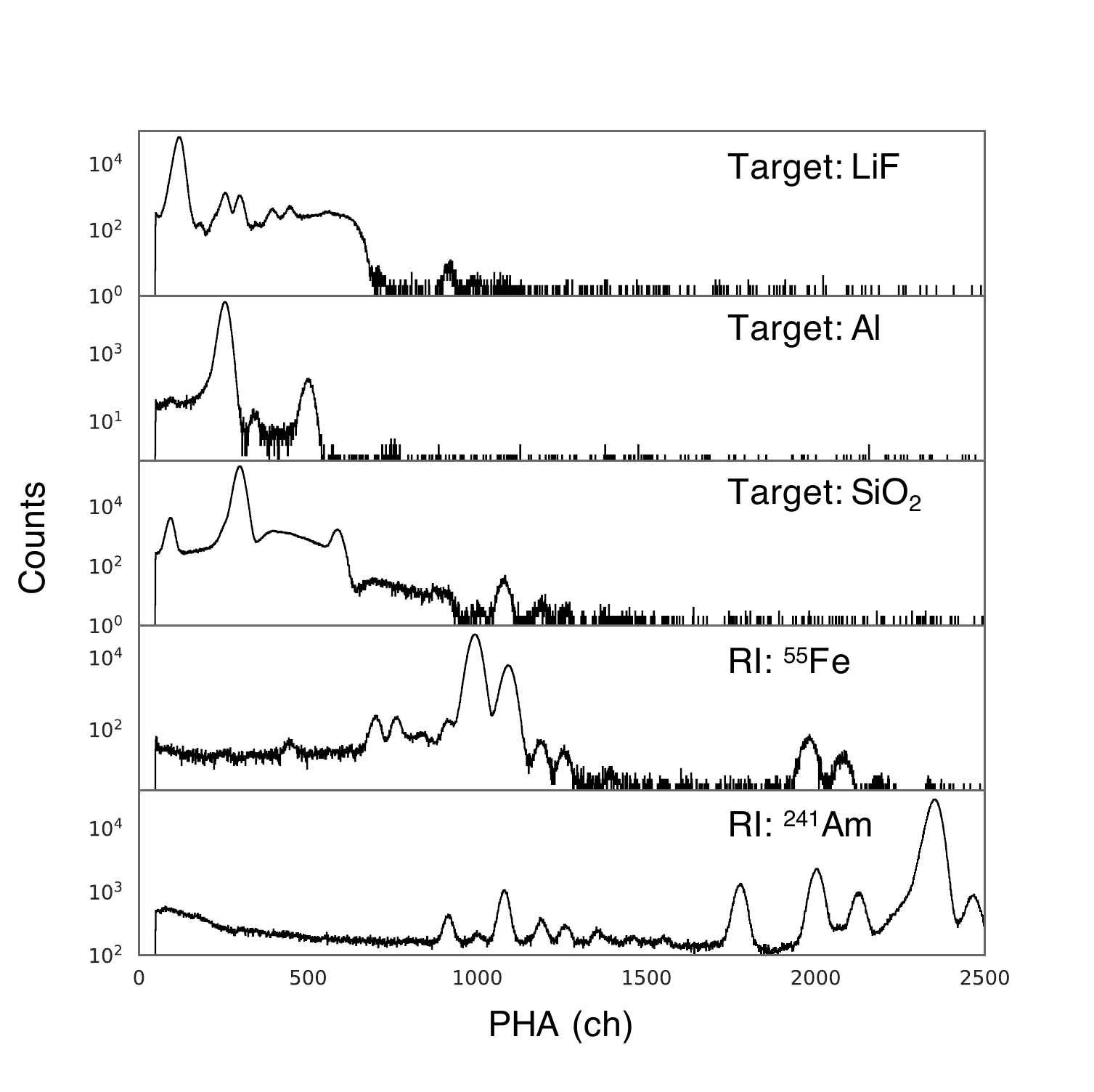}
 \caption[]{\label{fig:X-raySpectra} Spectra taken in the on-ground calibration
 using multi-color X-ray generator\cite{YONEYAMA2021164676}, in which the secondary
 targets as well as radioisotopes can be changed without breaking vacuum. Top three
 and bottom two show spectra using the secondary targets and the radioisotopes in
 the figure, respectively. Background lines from the materials contained in this
 system and pile-up lines are also seen.}
 \end{minipage}
 \hspace{0.02\hsize}
  \begin{minipage}{0.49\hsize}
   \centering 
   \vspace{3ex}
   \includegraphics[width=0.9\textwidth]{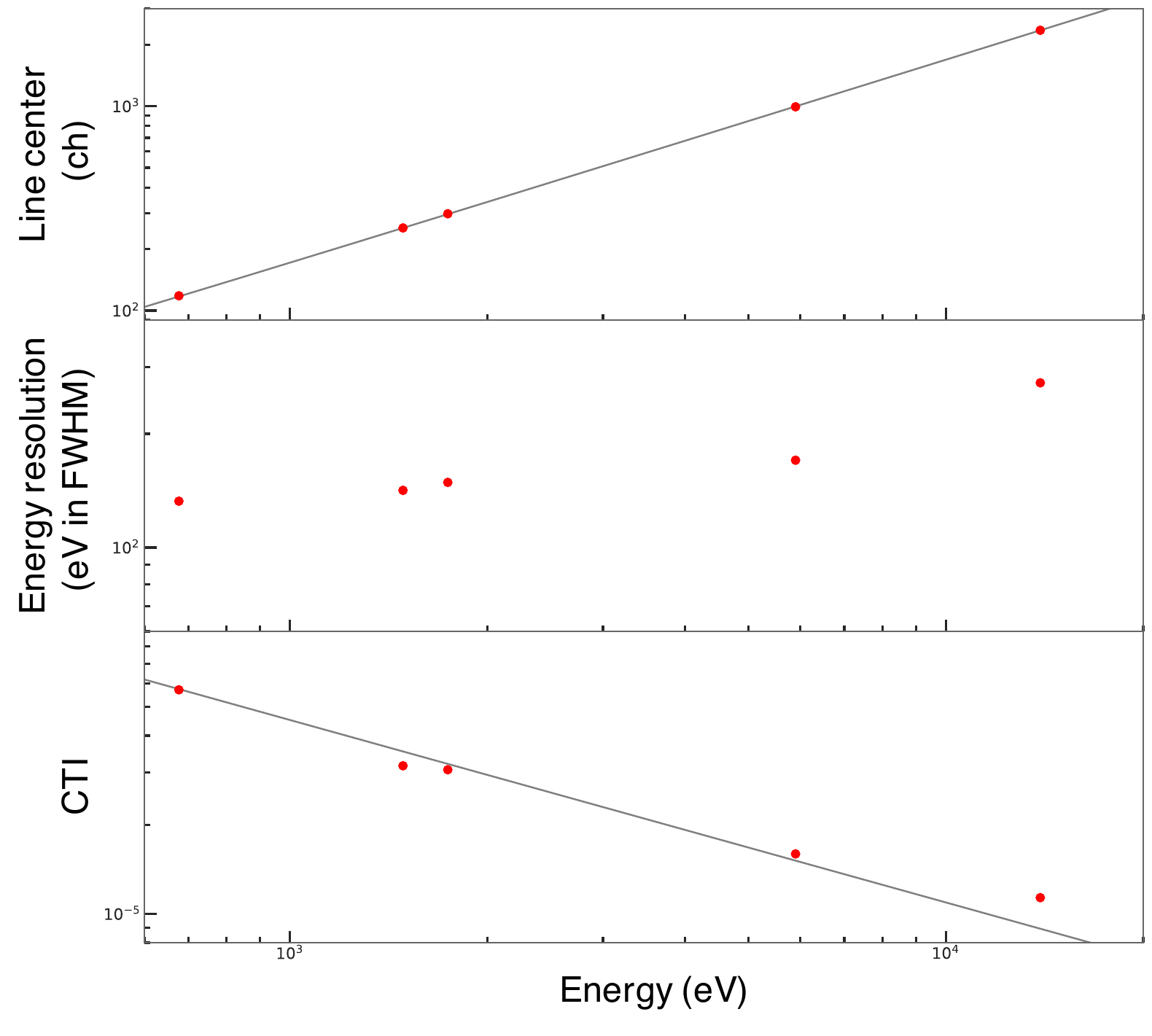}
 \caption[]{\label{fig:energyDependence} Energy dependence of line center (top),
 energy resolution (middle), and CTI (bottom). The values of the line center and
 energy resolution are derived from the single Gaussian fit to the spectra in
 Fig.~\ref{fig:X-raySpectra}. The CTI values in the bottom plot are specifically
 $c_{0\mathrm{F_{I}}}$ described in Ref.\citenum{KANEMARU2020164646}. The best-fit
   linear function and power function are superimposed in the top and bottom plot,
   respectively. }
 \end{minipage}
\end{figure}

We conducted the SXI on-ground calibration in two stages. In the first stage, the
flight CCDs were installed in a camera system built in our laboratory and their
spectroscopic performances were measured with multi-color X-ray data. The details of
the system and the multi-color X-ray generator were described in
Ref.\citenum{YONEYAMA2021164676}. Fig.~\ref{fig:X-raySpectra} shows X-ray spectra
taken in this system. The point in the first stage is that the noise environment and
overall gain in this system are different from those in the flight configuration
even if the degree of the differences would not be significant. Thus, the
calibration purpose in this stage is to derive calibration parameters that are
basically independent from the differences, and the main focus is to measure the
energy dependence of the line center, energy resolution, CTI, and so on, as shown in
Fig.~\ref{fig:energyDependence}. In the second stage, the calibration was performed
in the full flight configuration except for replacement of the camera bonnet in
order to make it vacuum-tight. An $^{55}$Fe source was attached in the non-flight
vacuum-tight bonnet and the entire areas of the four CCDs were irradiated with
X-rays from the source. The point in the second stage is that the data is available
only from the $^{55}$Fe source, not from multi-color X-ray sources. Thus, the
calibration purpose in this stage is to drive absolute calibration parameters at the
energies of Manganese lines. Combining the results from the first and second stage
measurements, we fully obtain the calibration parameters. The reason why we took
this two-stage way in our calibration is two-fold. One is from a safety reason. We
avoided unexpected risk that would have damaged flight electronics by using the
multi-color X-ray generator, which requires high voltage, onto the flight camera
body. The other is from a scheduling point of view. In this way, the flight camera
fabrication can be proceeded in parallel to the first stage of the calibration,
which was quite beneficial in the SXI development this time.

The on-ground calibration was performed with the CCD temperature of
$-110$~$^{\circ}$C, which is the initial operating temperature in orbit. On the
other hand, the thermal vacuum test was performed with the CCD temperature of
$-120$~$^{\circ}$C in order to verify the cooling capability of the SXI system.

\begin{figure} [ht]
  \centering \includegraphics[width=\textwidth]{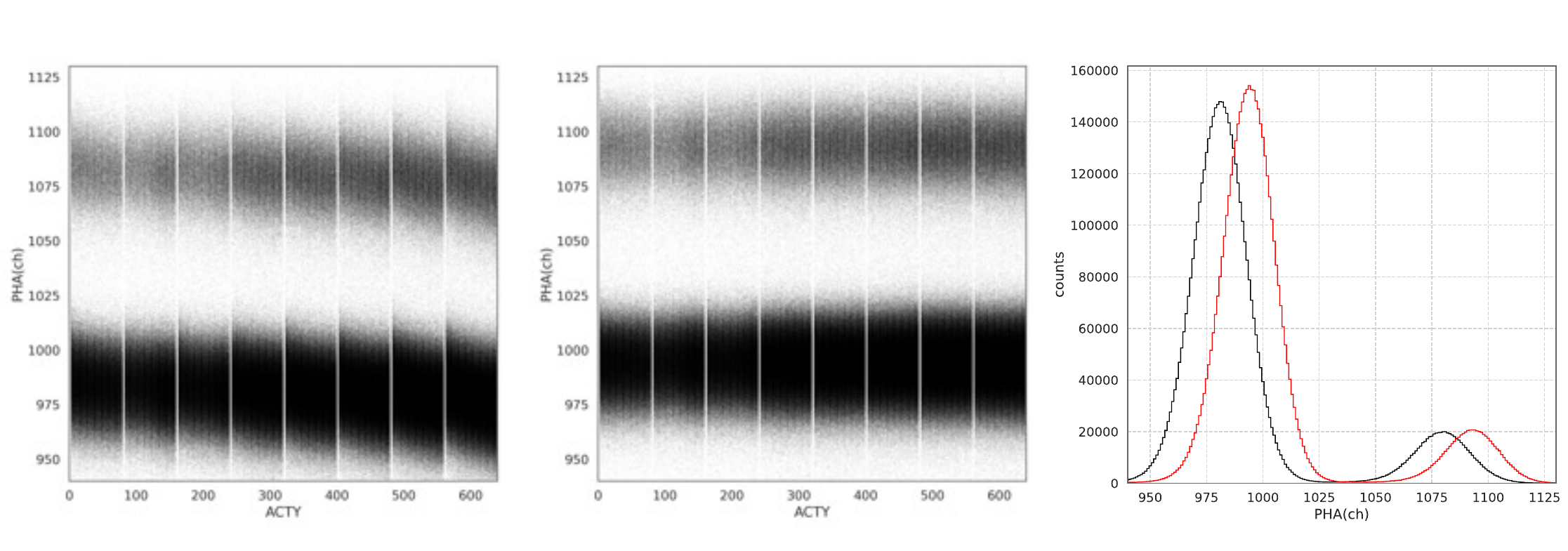}
  \caption[]{\label{fig:StackingPlot} Pulse heights of grade 0 events in an energy
  band which includes Mn-K$\alpha$ and Mn-K$\beta$ lines as a function of row number
  before (left) and after (middle) the CTI correction. (right) Spectrum before
  (black) and after (red) the CTI correction.}
\end{figure}

Fig.~\ref{fig:StackingPlot} shows pulse heights of grade 0 events in an energy band
which includes Mn-K$\alpha$ and Mn-K$\beta$ lines as a function of row number, which
is namely half of the number of transfer because of the on-chip binning, before and
after the CTI correction. Spectra before and after the CTI corrections also
shown. The data is taken in the second stage of the calibration. Parameters of the
CTI correction are derived following Ref.\citenum{KANEMARU2020164646}. Before the
CTI correction, the pulse height generally decreases as a function of row number
with periodic recovery at the charge injection (CI) rows, which are placed every 80
rows. This saw-tooth shape is the characteristics of the operation with the CI
technique~\cite{2009PASJ...61S...9U}. After the CTI correction, the saw-tooth shape
disappears and the pulse height does not depend on the row number as is expected in
the case without charge loss during transfer. It is also clear that the CTI
correction makes the peak and width of the spectrum higher and narrower,
respectively.


\begin{table}[ht]
 \caption{Summary of energy resolution measured with Mn-K$\alpha$ lines in the unit
 of eV in FWHM} \label{tab:SummaryOfDeltaE}
 \begin{center}       
  \begin{tabular}{|l|l|l|l|}
   \hline
   segment & entire region & cal.\ source region & on-axis region  \\
   \hline
   CCD1AB & 178.9$\pm$0.1 &  &  \\
   CCD1CD & 176.9$\pm$0.1 & 182.6$\pm$0.5 &  \\
   CCD2AB & 163.4$\pm$0.1 & 163.8$\pm$0.4 &  \\
   CCD2CD & 162.5$\pm$0.1 &  &  161.9$\pm$0.2\\
   CCD3AB & 166.0$\pm$0.1 & 168.1$\pm$0.3 &  \\
   CCD3CD & 168.4$\pm$0.1 &  &  \\
   CCD4AB & 164.5$\pm$0.1 &  &  \\
   CCD4CD & 168.9$\pm$0.1 & 173.2$\pm$0.4 &  \\
   \hline
  \end{tabular}
 \end{center}
\end{table}

Tab.~\ref{tab:SummaryOfDeltaE} summarizes the energy resolution of the CTI-corrected
spectra extracted from various regions. Each Mn-K$\alpha$ line is fitted with a
single Gaussian and the FWHM derived in the unit of eV are written. The ``cal.\ source
region'' and ``on-axis region'' are indicated by the green circles and the green box
in Fig.~\ref{fig:TV-img-spectrum} left, respectively. The spectrum taken from the
``entire region of the segment CCD2CD'' is actually shown in
Fig.~\ref{fig:StackingPlot}. At a given segment, the energy resolution of the cal.\
source region is always worse compared to that of the entire region because the
cal.\ source region is located at the farthest point from the read-out nodes. The
energy resolution of the cal.\ source region in CCD2AB shown here (163.8$\pm$0.4~eV)
is better than that in the thermal vacuum test (177.5$\pm$0.6~eV) in spite of the
higher temperature in the on-ground calibration. This is simply because the
improvement due to the CTI correction is more than the degradation due to the higher
temperature, especially for the data taken at the farthest point from the read-out
nodes. We note that the energy resolution of the cal.\ source region in CCD2AB was
178.2$\pm$0.4~eV before the CTI correction. Among the values in
Tab.~\ref{tab:SummaryOfDeltaE}, the best one is obtained at the on-axis region,
which is as we intended when we determined the position of the CCDs. The mission
requirement in the value of the energy resolution at the beginning of life is 200~eV
or less. All the values in Tab.~\ref{tab:SummaryOfDeltaE} are well below 200~eV with
sufficient margin and we confirmed that our flight CCDs satisfy the mission
requirement.

\begin{figure} [ht]
  \centering
  \includegraphics[width=0.4\textwidth]{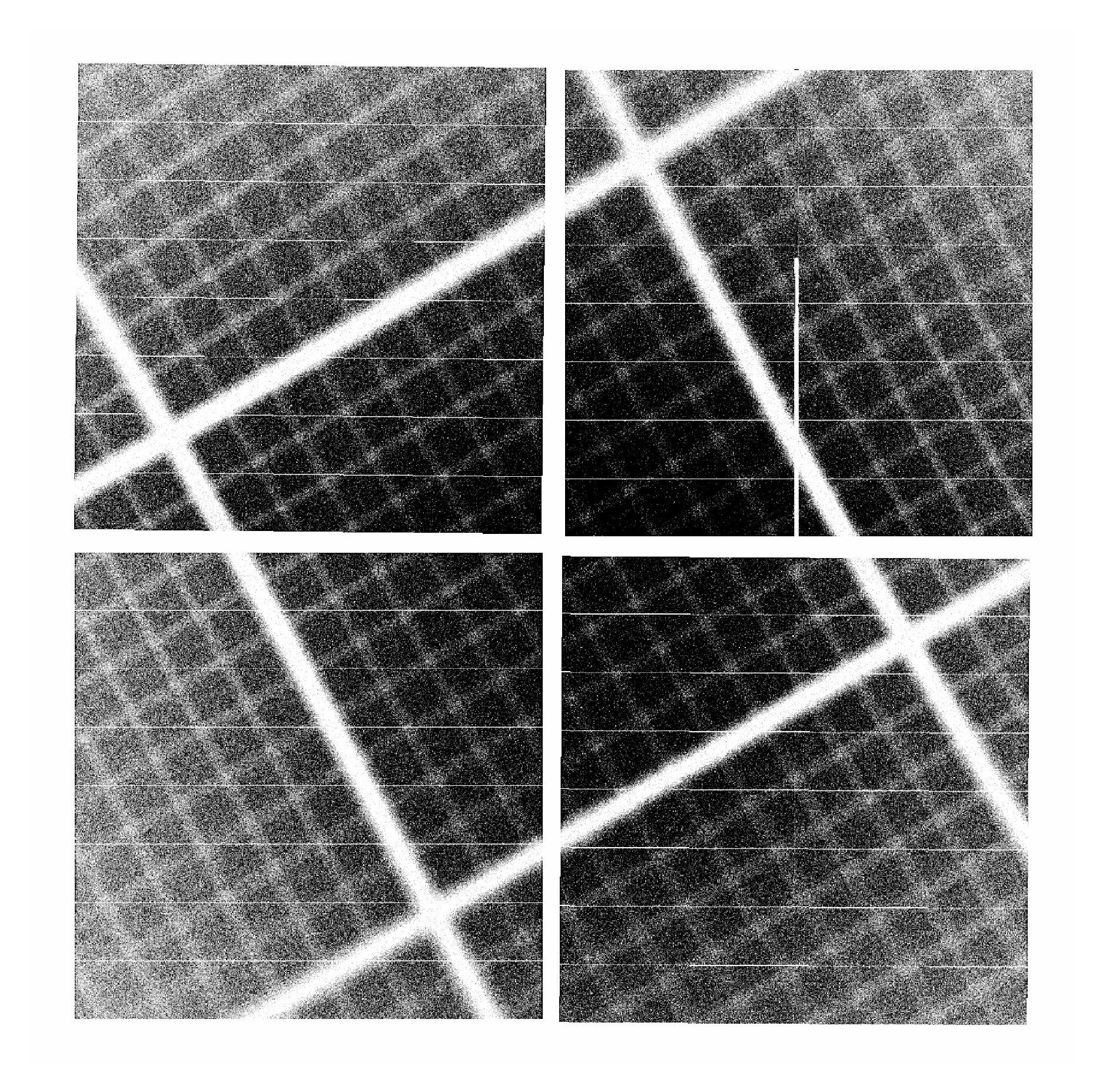}
  \caption[]{\label{fig:meshmap} Count map of X-ray events taken in the second stage
  of the on-ground calibration in a look-down view. A non-flight mesh frame
  was placed just above the CCDs in this experiment.}
\end{figure}

Fig.~\ref{fig:meshmap} shows a count map of X-ray events taken in the second stage
of the on-ground calibration. In this setup, a non-flight mesh frame was placed just above
the CCD. The mesh was removed when the vacuum-tight bonnet was replaced with the flight
bonnet before the installation of the SXI to the spacecraft body. We determined the
relative orientation of the CCDs aligning the shadows of the mesh bars. Gaps between
active pixel regions are then obtained to be
$45^{\prime\prime}$--$60^{\prime\prime}$. 

\section{SUMMARY}
\label{sec:summary}

We described the overview of Xtend, a soft X-ray imaging telescope onboard XRISM.
We completed the fabrication of the flight model of both SXI and XMA. Regarding the
SXI, a series of sub-system level performance verification tests was successfully
conducted. On-ground calibration measurements were also performed and the
spectroscopic performance of the SXI was confirmed to satisfy the mission
requirement.  Other calibration studies are ongoing. As of writing this paper, the
SXI system is installed into the spacecraft body and the XMA awaits the
installation.

\acknowledgments 

We are deeply saddened by the passing of our colleague and co-author, Kiyoshi
Hayashida. He launched the Xtend team and guided us until here as a principal
investigator. We will miss him as a colleague, as a friend, and above all, as a good
human being.

We acknowledge extensive supports from Hamamatsu Photonics K.K., Mitsubishi Heavy
Industries Ltd., and Sumitomo Heavy Industries Ltd.\ to develop CCD, SXI system, and
cooler component, respectively. This work was supported by JSPS KAKENHI Grant
Numbers JP21H01095, JP19K21884, JP20H01941, JP20H01947, JP21K03615, JP20KK0071,
JP18H01256, JP21H04972, JP25287042, JP22H01269, JP20H00175, JP21H04493, JP20K14491,
JP20J20685, JP21J00031, JP21K13963, JP21J10842, JP21K20372.


\bibliography{mybibfile} 
\bibliographystyle{spiebib} 

\end{document}